\documentclass[%
 reprint,%
 amssymb, amsmath,%
 aip,cha,%
]{revtex4-1}

\usepackage{docs}%
\usepackage{graphicx}
\usepackage{dcolumn}
\usepackage{bm}
\usepackage{units}
\expandafter\ifx\csname package@font\endcsname\relax\else
 \expandafter\expandafter
 \expandafter\usepackage
 \expandafter\expandafter
 \expandafter{\csname package@font\endcsname}%
\fi
\def\mathclap#1{\text{\hbox to 0pt{\hss$\mathsurround=0pt#1$\hss}}}

\newcommand{\msub}[1]{_{\text{#1}}}
\newcommand{\msup}[1]{^{\text{#1}}}

\DeclareFontFamily{U}{euc}{}
\DeclareFontShape{U}{euc}{m}{n}{<-6>eurm5<6-8>eurm7<8->eurm10}{}%
\DeclareSymbolFont{AMSc}{U}{euc}{m}{n} 
\DeclareMathSymbol{\umu}{\mathord}{AMSc}{"16}

\begin{document}

\preprint{AIP/123-QED}

\title[Extension of the measurement capabilities of the Quadrupole Resonator]{Extension of the Measurement Capabilities of the Quadrupole Resonator}
\thanks{Work supported by the German Doctoral Students program of the 
   Federal Ministry of Education and Research (BMBF)}

\author{T. Junginger}
 \altaffiliation[Also at ]{MPIK Heidelberg, Germany.}
\author{W. Weingarten}%
\affiliation{ 
CERN, Geneva, Switzerland
}%

\author{C. Welsch}
\affiliation{%
Cockcroft Institute, Warrington and University of Liverpool, United Kingdom
}%

\date{\today}

\begin{abstract}
The Quadrupole Resonator, designed to measure the surface resistance of superconducting samples at 400 MHz has been refurbished. The accuracy of its RF-DC compensation measurement technique is tested by an independent method. It is shown that the device enables also measurements at 800 and \unit[1200]{MHz} and is capable to probe the critical RF magnetic field. The electric and magnetic field configuration of the Quadrupole Resonator are dependent on the excited mode. It is shown how this can be used to distinguish between electric and magnetic losses.    
\end{abstract}

\maketitle

\section{Introduction}
The power dissipated in superconducting cavities is directly proportional to their surface resistance $R\msub{S}$, which shows a complex behavior on the external parameters: frequency $f$, temperature $T$, magnetic field $\vec{B}$ and electric field $\vec{E}$. In particular there is no widely accepted model which can describe the increase of the surface resistance with applied field. There is strong evidence that there are several different loss mechanisms, some only relevant if certain surface preparations are applied \cite{ciovatiHalbritter}. If not limited by a quench at a local defect, the maximum accelerating gradient of a superconducting cavity is set by the critical RF magnetic field. Its exact value and correlation to the surface properties of the material are not fully understood yet. Surface resistance and critical RF field can be directly measured in a superconducting cavity. However for the former its value obtained is the average $R\msub{S}$ over the whole surface. A convenient way to investigate the surface resistance and critical field of superconducting materials is to examine small samples, which can be manufactured at low cost, duplicated easily and used for further surface analyses. The Quadrupole Resonator \cite{Mahner:611593} was opted for measuring the surface resistance of superconducting niobium film samples at \unit[400]{MHz}, the technology and RF frequency chosen for the Large Hadron Collider (LHC) at CERN. The device is a four-wire transmission line half-wave resonator using a TE$_{21}$-like mode. The samples are thermally decoupled from the host cavity and their surface resistance is derived by a calorimetric RF-DC compensation technique. 

In this paper the extension of the Quadrupole Resonator to additionally cover the frequencies of 800 and \unit[1200]{MHz} and to probe the critical RF magnetic field of the samples is presented. It is shown how the frequency dependent field configuration on the sample surface can be used to distinguish between losses caused by the RF electric and magnetic field.
\begin{figure}[htbp]
	\centering
\includegraphics[width=0.95\columnwidth]{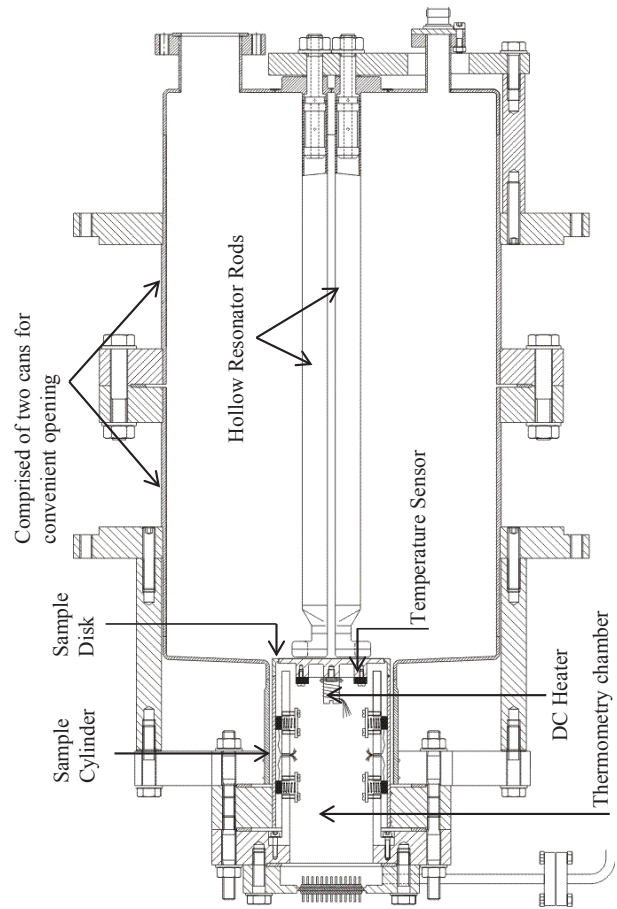}
	\caption{Technical drawing of the Quadrupole Resonator with attached sample and thermometry chamber housing a DC heater and temperature sensors \cite{Mahner:611593}.}
	\label{fig:Design2}
\end{figure} 

\section{Excitation at multiple frequencies}

The Quadrupole Resonator is a four-wire transmission line half-wave resonator. It was designed for excitation in a TE$_{210}$-like mode at \unit[400]{MHz}. The geometry also allows for excitation at multiple integers of \unit[400]{MHz} (TE$_{211}$, TE$_{212}$-like...). In the following it will be discussed whether these modes are also suited for surface resistance measurements on the attached samples.

 In the Quadrupole Resonator the cover plate of a cylinder attached to the cavity in a coaxial structure serves as the sample, see Fig.\ref{fig:Design2}. For the Quadrupole mode at \unit[400]{MHz} this design yields exponentially decaying RF fields between the outer wall of the sample cylinder and the host cavity. Therefore, the power dissipated inside this \unit[1]{mm} gap and especially at the end flange and joint of the sample cylinder is negligible. Additionally to the \unit[400]{MHz} design mode the fields are also exponentially decaying for all other excitable quadrupole (TE$_{21}$-like) modes up to \unit[2.0]{GHz}, as can be shown by analytical calculations \cite{Meinke:107091}. In principal five quadrupole modes could be excited and used for RF measurements. At CERN equipment for 400, 800 and \unit[1200]{MHz} is available for the test stand. 

The Quadrupole Resonator consists of two \unit[2]{mm} thick niobium cans for convenient handling and cleaning of the device, see Fig. \ref{fig:Design2}. These cans are flanged to each other in the middle of the resonator, where the screening current on the cavity surface vanishes for the modes at 400 and \unit[1200]{MHz}. For the \unit[800]{MHz} mode the screening current has a maximum at this position. However, since the field is strongly concentrated around the rods in the middle of the resonator, excitation and measurements at \unit[800]{MHz} are not perturbed by losses at this flange.  In fact the magnetic field at the flange between the upper and the lower can is only \unit[0.5]{\%} of the maximum field on the sample as has been calculated by Microwave Studio \textsuperscript{\textregistered} (MWS) \cite{cstmws}. The Quadrupole Resonator is equipped with two identical strongly overcoupled antennas. One serves as the input, the other as the output. Due to this configuration almost the whole power transmitted to the cavity is coupled out and only about \unit[1]{\%} is dissipated in the cavity walls and on the sample surface. The system acts like a narrow band filter with minor losses. At \unit[800]{MHz} it remains in this strongly overcoupled state up to the highest field level reached, which is about \unit[40]{mT} for this frequency. No deviation of the coupling strength at higher field levels was ever observed if sample and cavity remained in a superconducting state. Therefore the design comprising the two cans does not perturb the measurements at \unit[800]{MHz}.

At \unit[1200]{MHz} measurements are more cumbersome due to \unit[69]{Hz} oscillations of the resonator rods. It could be shown that these vibrations are excited by helium bubbles forming in the resonator rods, because they are suppressed, but not completely avoided, when measurements are performed inside a superfluid helium bath. 

\section{Low field surface resistance}
\begin{figure}[tb]
   \centering
    \includegraphics[width=\columnwidth]{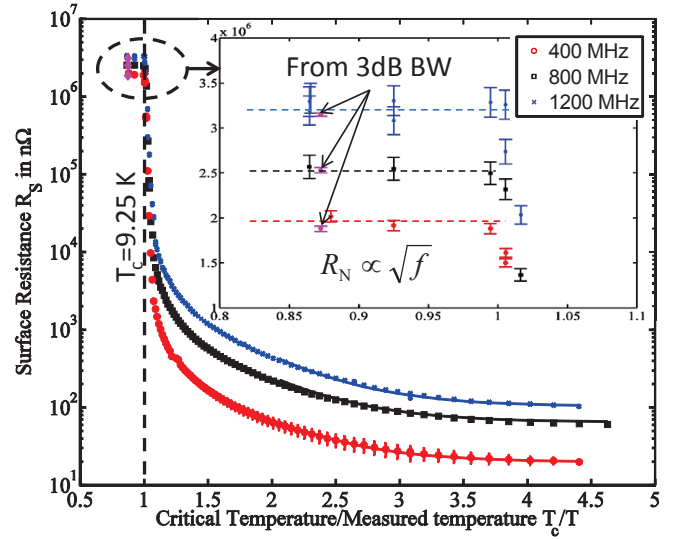}
   \caption[Low field surface resistance of a reactor grade bulk niobium sample]{Surface resistance of a reactor grade bulk niobium sample for three frequencies as a function of temperature, measured at a magnetic field of about \unit[15]{mT}. The solid lines show least squares fits to BCS theory. In the normal conducting regime above $T\msub{c}$ the surface resistance was also derived from \unit[3]{dB} bandwidth measurements.}
   \label{figure:NbRT}
\end{figure}

For magnetic fields below \unit[15]{mT} the surface resistance $R\msub{S}$ is assumed to be independent of the field strength and thus can be written as a sum of BCS and residual surface resistance, 
\begin{equation}
R\msub{S}=R\msub{BCS}(f,T)+R\msub{Res}(f).
\end{equation} 
Figure \ref{figure:NbRT} displays $R\msub{S}$ for an applied RF magnetic field of approx.\,\unit[15]{mT} in the temperature range between 2 and \unit[10.6]{K} for a reactor grade bulk niobium sample, which was prepared by buffered chemical polishing (BCP).  

In the normal conducting regime above $T\msub{c}$, the surface resistance depends only slightly on temperature as can be seen from the flat curves in this area. For these temperatures the normal conducting surface resistance of the sample $R\msub{N}$ can also be derived by a non-calorimetric \unit[3]{dB} bandwidth method \cite{3dBBW}. The measurement is only possible when the sample is normal conducting and the coupling changes from strongly overcoupled to strongly undercoupled. This means that in the latter case the majority of the power coupled in is dissipated on the sample surface instead of being coupled out. The loaded quality factor $Q\msub{L}$ consists of the quality factors of the host cavity $Q\msub{C}$, the sample $Q\msub{Sample}$ and the two couplers combined $Q\msub{ext}$
\begin{equation}
\frac{1}{Q\msub{L}}=\frac{1}{Q\msub{ext}}+\frac{1}{Q\msub{C}}+\frac{1}{Q\msub{Sample}}.
\label{eq:QL}
\end{equation}
In the superconducting state $Q\msub{L}$ was measured to be about $10^6$, dependent on frequency. The quality factor of the sample $Q\msub{Sample}$ can be derived from the calorimetric measurement:
\begin{equation}
	R{\msub{S}}=\frac{G\msub{Sample}}{Q\msub{Sample}},
	\label{eq:Q0Sample}
\end{equation}
where $G\msub{Sample}$ is the geometry factor of the sample. It relates the losses on the sample surface to the stored energy in the cavity $U$: 
\begin{equation}
	G\msub{Sample}= \mu_0^2\frac{2\omega U}{\int\limits_{\mathclap{\mathrm{Sample}}}{{B^2}\mathrm{d}S}}.
\label{eq:GSample}
\end{equation}
At \unit[2]{K} $R\msub{S}$ is several tens of nanoohms, corresponding to $Q\msub{Sample}$ values of several times $10^9$. The host cavity is made of the same material as the sample. Therefore, its surface resistance is considered to be almost identical. From this assumption and the field configuration of the Quadrupole Resonator (about ten percent of the power is dissipated on the sample surface) one can estimate $Q\msub{C}$ to be about ten times lower than $Q\msub{Sample}$. This is still two orders of magnitude higher than $Q\msub{L}$, allowing to simplify Eq.\,\ref{eq:QL} to 
\begin{equation}
\frac{1}{Q\msub{L}}=\frac{1}{Q\msub{ext}}.
\label{eq:QL2}
\end{equation}
If the sample is normal conducting the system becomes undercoupled. 
The host cavity remains superconducting, since it is thermally decoupled from the sample. Thus, $1/Q\msub{C}$ remains negligible and Eq.\,\eqref{eq:QL} reads
\begin{equation}
\frac{1}{Q\msub{L}}=\frac{1}{Q\msub{ext}}+\frac{1}{Q\msub{Sample}}.
\label{eq:QL3}
\end{equation}
$Q\msub{Sample}$ can now be calculated with the value of $Q\msub{ext}$ obtained from the measurement in the superconducting
state. From $Q\msub{Sample}$ $R\msub{S}$ is derived using Eq.\,\eqref{eq:Q0Sample}.  
%
%
%
%
%
%
The results from these non-calorimetric \unit[3]{dB} bandwidth measurements agree within \unit[4]{\%} with the results obtained from the calorimetric measurements, see Fig.\,\ref{figure:NbRT}. The surface resistance in this normal conducting regime is found to be proportional to $\sqrt{f}$ as expected in case of normal skin effect. These results confirm the validity of the calorimetric approach and therefore also give confidence in the measurements at lower temperatures. 

The solid lines in Fig.\,\ref{figure:NbRT} show the predictions from least squares fits to BCS theory performed with Win Super Fit \cite{Ciovati_Win_super_Fit,ciovatiphd}. The program uses the Levenberg-Marquardt algorithm \cite{Levenberg,Marquardt} for $\chi^2$ minimization and is based on the widely used Halbritter code for the calculation of the surface resistance \cite{Halbritter_Fortran}. For the data presented here the superconducting energy gap $\Delta$ and the residual resistance $R\msub{Res}$ were varied to minimize $\chi^2$.   
The derived values for both samples are given in Tab.\,\ref{tab:low_field_bulk}. Other input parameters, which were set constant in the program are the critical temperature $T\msub{c}$=\unit[9.25]{K} and the mean free path $l$=\unit[110]{nm} derived from penetration depth measurements, while the BCS coherence length $\xi_0$=\unit[39]{nm} and the London penetration depth $\lambda_L$=\unit[33]{nm} were taken from the literature \cite{MattisBardeenTheory}. The energy gap $\Delta$ is found consistent for the three frequencies. The values are also consistent with theory and other measurements \cite{MattisBardeenTheory}. The obtained residual resistance $R\msub{Res}$ is as expected for reactor grade niobium material \cite{SafaCryoEng}. It is obviously dependent on frequency, as has also been pointed out in other publications, where $R\msub{Res}$ was derived with a multimode cavity \cite{Halama} and for a large batch of elliptical cavities of same surface treatment but different resonant frequency \cite{CiovatiResidual}. The former approach is limited by the different field configurations for each mode and the latter can give only a statistical result, since every cavity surface is different. In contrast the results from the Quadrupole Resonator are obtained for the same sample under almost identical magnetic field configuration on the surface. 
%
 %
%
\begin{table}[b]
   \centering
\caption[Material parameters derived from low field surface resistance measurements]{Material parameters derived from low field surface resistance measurements below \unit[4.5]{K}}
\begin{tabular}{lcc}
  \hline
$f$ in MHz & $\Delta /\textrm{k}\msub{B}$ in K & $R\msub{Res}$ in n$\Omega$  \\ 
\hline
400 & 18.2$\pm$0.5 & 19.8$\pm$0.8 \\  
800 & 18.4$\pm$0.4 & 62.8$\pm$0.6  \\ 
1200 & 18.5$\pm$0.2 & 99.8$\pm$0.7 \\
\hline 
\end{tabular}
\label{tab:low_field_bulk}
\end{table} 
%
%
%
\section{Electric and Magnetic Field Configuration}
\label{sec:Electric and Magnetic Field Configuration}
Cylindrical cavities operated in a TE mode are often used for material characterization. These cavities expose the samples attached only to an RF magnetic field. The Quadrupole Resonator with its different field configuration exposes the samples to electric and magnetic fields simultaneously. The electric field $\vec{E}$ on the Quadrupole Resonator sample surface scales linearly with frequency for a given magnetic field $\vec{B}$, as required by the law of induction when applied to the geometry in between the crooked endings of the rods and the sample. For a peak magnetic field  $B\msub{p}$=\unit[10]{mT}, the peak electric field is $E\msub{p}=$\unit[0.52,\,1.04,\,1.56]{MV/m} for \unit[400, 800 and 1200]{MHz} respectively as has been calculated using Microwave Studio. These values are small compared to $E\msub{p}$-field levels on elliptical cavities, but the area of high electric field is larger. In elliptical cavities the surface electric field is mainly concentrated around the iris of the cavity. In the Quadrupole Resonator it is approximately spread over the same area on the sample surface as the magnetic field. The fact that the ratio of the mean values $E\msub{mean}$/$B\msub{mean}$ for elliptical cavities and the Quadrupole Resonator are comparable is a valuable feature if real accelerator cavity surfaces are to be studied. 

In the following the implications of the field dependent ratio $E/B$ on the interpretation on measurement results is discussed. The Quadrupole Resonator measures the power dissipated on the surface of the attached sample by a calorimetric RF-DC compensation measurement consisting of two steps: 
\begin{enumerate}
	\item{The temperature of interest is set by applying a current to the resistor on the back side of the sample. The power dissipated $P\msub{DC,1}$ is derived from measuring the voltage across the resistor.}
	\item{The RF is switched on and the current applied to the resistor is lowered to keep the sample temperature and the total power dissipated constant.}
\end{enumerate}
The power dissipated by RF, $P_{\mathrm{RF}}$ is the difference between the DC power applied without RF, $P_{\mathrm{DC1}}$ and the DC power applied with RF, $P_{\mathrm{DC2}}$.
In general these losses are caused by the RF magnetic and electric fields with the two contributions being additive,    
\begin{equation}
P\msub{RF}=P\msub{DC1}-P\msub{DC2}=\frac{1}{2\mu_0^2} \int\limits_{\mathclap{\mathrm{Sample}}}R_{\mathrm{S}}\vec{B}^2\mathrm{d}S
+\frac{\varepsilon_0}{2\mu_0} \int\limits_{\mathclap{\mathrm{Sample}}} R_{\mathrm{S}}\msup{E}\vec{E}^2\mathrm{d}S,
\label{eq:PRF}
\end{equation}
where $R_{\mathrm{S}}\msup{E}$ is the electrical surface resistance. Usually, when the surface resistance of superconducting cavities is investigated, the losses are assumed to be caused by the RF magnetic field $B$, since the contribution from the electric field $E$ is negligible, even for normal conducting metals \cite{HR_electric_surface_impedance}. However, for oxidized surfaces additional loss mechanisms need to be taken into account.
To relate electric and magnetic losses to each other the constant $c$ is introduced
\begin{equation}
c=\frac{G\msub{Sample}\msup{E}}{G\msub{Sample}}
\end{equation}
This ratio between the magnetic and the electric geometry factor scales quadratically with frequency. This follows directly from the law of induction for the Quadrupole Resonator geometry and has been verified with an agreement better than 1\% using MWS. This allows to normalize $c$ to \unit[400]{MHz}
\begin{equation}
c(f)=\frac{G\msup{E}\msub{Sample}(f)}{G\msub{Sample}(f)}=c\msub{400}\frac{(400\,\textrm{MHz})^2}{f^2}.
\label{eq:c3norm}
\end{equation}
Microwave Studio was used to calculate $c_{400}$=53.5.
This implies that a power dissipated by the RF field on the sample surface $P\msub{RF}$, corresponding to a magnetic surface resistance of \unit[1]{n$\Omega$}, is equivalent to an electric surface resistance of \unit[53.5]{n$\Omega$} at \unit[400]{MHz}, while a measured RF heating corresponding to $R\msub{S}$=\unit[1]{n$\Omega$} at \unit[800]{MHz} is only equivalent to $R\msub{S}\msup{E}$=\unit[13.4]{n$\Omega$} .     

The Quadrupole Resonator does not allow to measure the magnetic and the electric losses independently. One has to measure the complete losses and then the interpret the data. For example the surface resistance as shown in Tab. \ref{tab:low_field_bulk} has been calculated assuming the residual resistance is completely caused by the RF magnetic field. Assuming that it is caused by the electric field, the values are $R\msub{S}\msup{E}$=\unit[(1060$\pm$40)]{n$\Omega$} at \unit[400]{MHz}, $R\msub{S}\msup{E}$=\unit[(840$\pm$8)]{n$\Omega$} at \unit[800]{MHz} and $R\msub{S}\msup{E}$=\unit[(593$\pm$4)]{n$\Omega$} at \unit[1200]{MHz}. This would imply an unphysical higher electric surface resistance at lower frequency, which allows to conclude that $R\msub{Res}$ is at least mainly caused by the magnetic field. The frequency dependent ratio between $E\msub{p}$ and $B\msub{p}$ enabled to reveal electric losses due to interface tunnel exchange to be the cause for a field dependent surface resistance at RF electric fields of a few MV/m on oxidized granular surfaces, see...   

\section{Maximum RF Field}
%
In the following it will be shown that the Quadrupole Resonator, designed for surface resistance measurements, is also suited to probe the intrinsic maximum RF magnetic field $B\msub{max,RF}$ of the samples.
%
A quench is detected from a sudden drop of the transmitted power by several orders of magnitude. One can easily determine, without further diagnostics, if it happened on the host cavity or on the sample by measuring the sample temperature at the moment the quench occurs. If the temperature rises above the critical temperature $T\msub{c}$ it was on the sample, otherwise it must have been on the host cavity. 

\begin{figure}[tb]
   \centering
 \includegraphics[width=0.95\columnwidth]{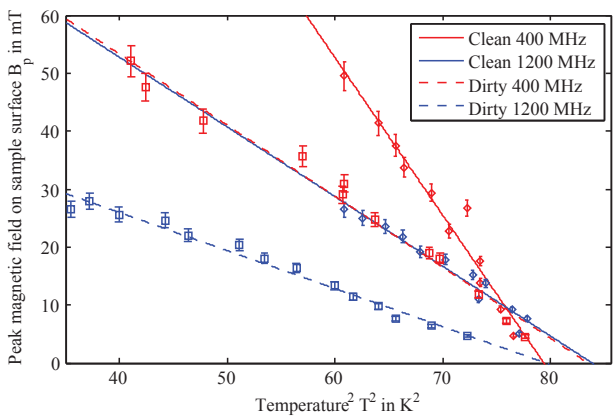}
   \caption{Quench field $B\msub{max,RF}$ of the bulk niobium sample in clean and dirty condition, measured in CW.}
  \label{figure:Quench_CW}
\end{figure}

The critical field under RF exposure has been investigated using pulses just long enough that the stored energy in the cavity reaches steady state (pulse length approx. \unit[2]{ms}) and also in continuous wave (CW) operation. Different field levels and dependencies on frequency have been found for each case. In the analysis of the CW measurements it is assumed that $B\msub{max,RF}$ has the same dependence on temperature as the critical thermodynamic field $B\msub{c}$ and can therefore be written as
\begin{equation}
B\msub{max,RF}(T)=B\msub{max,RF}(0)\left(1-\left(\frac{T}{T\msub{c}}\right)^2\right).
\end{equation}
In order to measure the critical field in continuous wave (CW), first the magnetic field on the sample surface $B\msub{p}$ is set to a fixed level. Then the sample temperature is slowly raised until the quench occurs. Usually a sudden temperature rise above $T\msub{c}$ is observed at the moment the quench occurs.
%
%
%
%
%
%
%
When measured in CW the quench field is dependent on frequency and surface properties. In early tests a bulk niobium sample quenched at relatively low field levels due to a local defect. A second etching (BCP 100 $\umu$m) yielded higher field levels for the clean sample, see Fig.\,\ref{figure:Quench_CW}. The fact that $B\msub{p}$ vs. $T^2$ gives a straight line is an indication that an intrinsic superconducting field limitation is found for all curves. This can be explained by a local defect heating its surrounding area. When the temperature in the vicinity of the defect exceeds the field dependent critical temperature the quench occurs. 
%
%
%
%
%
%
%

Fitting a straight line to the data displayed in Fig.\ref{figure:Quench_CW} allows to derive $T\msub{c}$ and $B\msub{max,RF}(0)$. The intersection of each line with the x-axis gives $T\msub{c}^2$ and the slope $B\msub{max,RF}(0)/T\msub{c}^2$. The parameter values found for all curves are listed in Tab.\,\ref{tab:Fit_Parameter_Hcrit}.
\begin{table}[b]
   \centering
   \caption{Critical RF field of the bulk niobium sample}
   \begin{tabular}{lcccc}
       \toprule
       Measurement &Sample    &  $f$     &  $B\msub{max,RF}(0)$ & $T\msub{c}$  \\
       technique   &condition &	in MHz	 &	in mT               & in K \\		
       \hline
    CW & dirty & 400 & 102$\pm$8 & 9.14$\pm$0.14 \\
    CW & dirty & 1200 & 52$\pm$3 & 8.92$\pm$0.14 \\
    CW & clean & 400 & 216$\pm$29 & 8.91$\pm$0.09 \\
    CW & clean & 1200 & 101$\pm$13 & 9.16$\pm$0.12 \\
    Pulsed & clean & 400 & 243$\pm$15 & 9.11$\pm$0.06 \\
       \hline
   \end{tabular}
   \label{tab:Fit_Parameter_Hcrit}
\end{table}
The quench field in CW is about twice as high at the lower frequency. This cannot be explained by pure thermal breakdown caused by a local normal conducting defect \cite{Wolfgang_USCERN}.
In this case the quench field would scale with $f^{-0.25}$ assuming normal skin effect. This would imply that the quench field at \unit[400]{MHz} is only 1.3 times higher than at \unit[1200]{MHz}. This prediction is clearly in contradiction to the measurement results presented here. It can also be excluded that the quench is a complete magnetic effect. In this case the same maximum field should be reached independent of frequency and duty cycle. The RF breakdown can be explained by taking both magnetic and thermal effects into account. That both effects are relevant for RF breakdown was also recently found by Eremeev et al. \cite{Eeremeev2011}. 

%
%
Measured in pulsed operation a lower critical temperature compared to the value derived from the low field surface resistance measurements was obtained. This is due to the fact that the position of highest field value is located about \unit[1.5]{mm} closer to the heater than the position of the temperature sensors, which consequently indicates a lower temperature than prevails at the region of maximum field. For the following comparison of the maximum RF field with theory the value of $T\msub{c}$=\unit[9.11]{K} derived from the pulsed measurement will be used.    
\begin{figure}[tb]
   \centering
    \includegraphics[width=0.95\columnwidth]{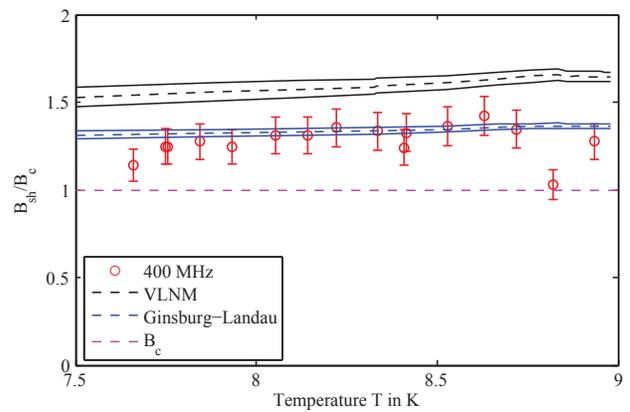}
   \caption{Critical field under RF (short pulses) of the clean bulk niobium sample.}
   \label{figure:Quench_Pulsed2}
\end{figure}

Figure\,\ref{figure:Quench_Pulsed2} shows the maximum RF field normalized to the thermodynamic critical field $B\msub{c}$=\unit[199]{mT} \cite{Leupold,Ferreira} as a function of temperature. Here  it can be seen that the maximum RF field systematically exceeds $B\msub{c}$. Critical RF fields above $B\msub{c}$ have also been measured in several other publications \cite{Hays97,VallesSRF2009,Valles2010}. Their results can be explained by a superheating field $B\msub{sh}$ either derived from considering a metastable state preventing flux entry in the superconductor by a surface barrier \cite{Matricon1967241,PhysRevB.83.094505} or by considering a thermodynamic energy balance at the interface between the superconductor and the adjacent vacuum \cite{PhysRevLett.39.826,saitosrf2001,saitosrf2003}. Figure\,\ref{figure:Quench_Pulsed2} shows the predictions from the vortex line nucleation model (VLNM) \cite{saitosrf2003} and the approximate formulas from \cite{PhysRevB.83.094505} based on Ginsburg-Landau theory, therefore in the following named the Ginsburg-Landau model (GLM). The VLNM is based on a thermodynamic energy balance, while the GLM considers metastability. Both models relate the superheating field to the Ginsburg-Landau parameter $\kappa$ and the critical thermodynamic field $B\msub{c}$. The latter parameter was taken from literature, while $\kappa$ has been calculated from penetration depth measurements. The error bands are directly correlated to the uncertainties of $\kappa$. For low values of $\kappa$ (long mean free path) the Ginsburg-Landau model predicts $B\msub{sh}$ smaller than the VLNM. This is the case for the bulk niobium sample in clean condition, for which $B\msub{max,RF}$ was measured with values consistent with the Ginsburg-Landau model, see Fig.\,\ref{figure:Quench_Pulsed2}.
%
%
%
%
\begin{figure}[tb]
   \centering
    \includegraphics[width=0.95\columnwidth]{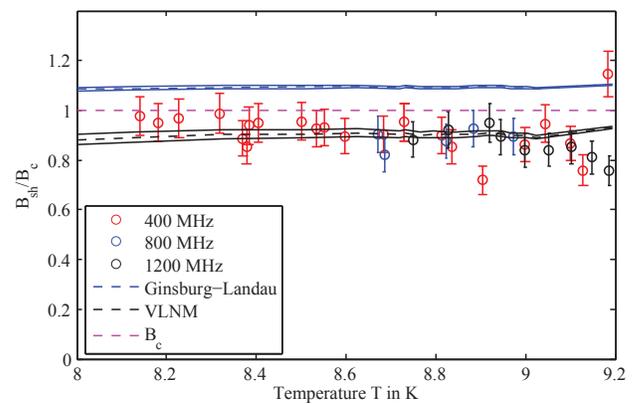}
   \caption{Critical field under RF (short pulses) of the niobium film sample.}
   \label{figure:Quench_Pulsed_NbCU}
\end{figure}  
For high values of $\kappa$ (low mean free path) the Ginsburg-Landau model predicts $B\msub{sh}$ higher than the VLNM. This condition was found for a niobium film sample, which was prepared by DC magnetron sputtering onto a copper substrate. 
%
The critical temperature of this sample was measured to be $T\msub{c}$=\unit[9.28]{$\pm$0.03} by the same approach as described above for the bulk niobium sample, while the Ginsburg-Landau parameter $\kappa$ was again derived from penetration depth measurements.   
%
%
The maximum RF field for this sample does not exceed $B\msub{c}$. The values are consistent with the VLNM, see Fig.\,\ref{figure:Quench_Pulsed_NbCU}. For all three frequencies the same value of $B\msub{max,RF}$ is found if sufficiently short pulses are used. This shows that the intrinsic critical RF field can be measured with the Quadrupole Resonator. 

From the analysis of the critical field of one sample of low and one of high purity it seems that the GLM and the VLNM can both limit the maximum field under RF in superconducting cavities. For both samples the lower barrier could not be exceeded.   
It can however not be excluded that the limitation set by the VLNM can be overcome for cleaner surfaces. It has been stated that for very high values $\kappa$ this model predicts unrealistic low values for the superheating field \cite{PhysRevB.78.224509}.

\section{Summary}
The Quadrupole Resonator has been refurbished and its measurement capabilities have been extended. The calorimetric results were thereby verified by an independent technique. It was shown that the device can be used for surface resistance measurements at multiple integers of its design frequency. The almost identical magnetic field configurations in combination with the mode dependent electrical field configuration enable unique possibilities to test theoretical surface resistance models. Additionally it was shown that the intrinsic critical RF magnetic field can be measured with the device. 

\section{Acknowledgment}
The authors would like to thank everybody who contributed to the refurbishment and operation of the Quadrupole Resonator. The work of S. Calatroni and S. Forel (preparation of samples) and the operators from the CERN cryogenics group is highly appreciated. We thank Ernst Haebel, now retired, for explaining to us the original idea and conception of the Quadrupole Resonator. The help of N. Schwerg performing the \unit[3]{dB} bandwidth measurements is appreciated. 
%

%

%
%
\end{document}